\documentclass[10pt,conference]{IEEEtran}

\usepackage{graphicx,epic,eepic,epsfig,amsmath,latexsym,amssymb,verbatim,subfigure,color}
\usepackage{theorem}

\def\squareforqed{\hbox{\rlap{$\sqcap$}$\sqcup$}}
\def\qed{\ifmmode\squareforqed\else{\unskip\nobreak\hfil
\penalty50\hskip1em\null\nobreak\hfil\squareforqed
\parfillskip=0pt\finalhyphendemerits=0\endgraf}\fi}
\def\endenv{\ifmmode\;\else{\unskip\nobreak\hfil
\penalty50\hskip1em\null\nobreak\hfil\;
\parfillskip=0pt\finalhyphendemerits=0\endgraf}\fi}

\mathchardef\ordinarycolon\mathcode`\:
\mathcode`\:=\string"8000
\def\vcentcolon{\mathrel{\mathop\ordinarycolon}}
\begingroup \catcode`\:=\active
  \lowercase{\endgroup
  \let :\vcentcolon
  }

\newcommand{\nc}{\newcommand}
\nc{\rnc}{\renewcommand}
\nc{\beq}{\begin{equation}}
\nc{\eeq}{{\end{equation}}}
\nc{\beqa}{\begin{eqnarray}}
\nc{\eeqa}{\end{eqnarray}}
\nc{\lbar}[1]{\overline{#1}}
\nc{\bra}[1]{\langle#1|}
\nc{\ket}[1]{|#1\rangle}
\nc{\ketbra}[2]{|#1\rangle\!\langle#2|}
\nc{\braket}[2]{\langle#1|#2\rangle}
\nc{\proj}[1]{| #1\rangle\!\langle #1 |}
\nc{\avg}[1]{\langle#1\rangle}
\nc{\Rank}{\operatorname{Rank}}
\nc{\smfrac}[2]{\mbox{$\frac{#1}{#2}$}}
\nc{\Tr}{\operatorname{Tr}}
\nc{\tr}{\operatorname{Tr}}
\nc{\id}{\operatorname{id}}
\nc{\1}{\openone}
\nc{\ox}{\otimes}
\nc{\dg}{\dagger}
\nc{\dn}{\downarrow}
\nc{\cA}{{\cal A}}
\nc{\cB}{{\cal B}}
\nc{\cC}{{\cal C}}
\nc{\cD}{{\cal D}}
\nc{\cE}{{\cal E}}
\nc{\cF}{{\cal F}}
\nc{\cG}{{\cal G}}
\nc{\cH}{{\cal H}}
\nc{\cI}{{\cal I}}
\nc{\cJ}{{\cal J}}
\nc{\cK}{{\cal K}}
\nc{\cL}{{\cal L}}
\nc{\cM}{{\cal M}}
\nc{\cN}{{\cal N}}
\nc{\cO}{{\cal O}}
\nc{\cP}{{\cal P}}
\nc{\cR}{{\cal R}}
\nc{\cS}{{\cal S}}
\nc{\cT}{{\cal T}}
\nc{\cX}{{\cal X}}
\nc{\cY}{{\cal Y}}
\nc{\cZ}{{\cal Z}}
\nc{\supp}{{\operatorname{supp}}}
\nc{\var}{\operatorname{var}}
\nc{\rar}{\rightarrow}
\nc{\lrar}{\longrightarrow}
\nc{\polylog}{\operatorname{polylog}}

\nc{\RR}{{{\mathbb R}}}
\nc{\CC}{{{\mathbb C}}}
\nc{\FF}{{{\mathbb F}}}
\nc{\NN}{{{\mathbb N}}}
\nc{\ZZ}{{{\mathbb Z}}}
\nc{\PP}{{{\mathbb P}}}
\nc{\QQ}{{{\mathbb Q}}}
\nc{\UU}{{{\mathbb U}}}
\nc{\EE}{{{\mathbb E}}}
\nc{\Icoh}{{I^c}}
\nc{\Qca}{{Q_{\rm ss}}}
\nc{\Qcaa}{{Q^{(1)}_{\rm ss}}}
\nc{\Dcaa}{{D^{(1)}_{{\rm ss}\rightarrow}}}
\nc{\Dca}{{D_{{\rm ss}\rightarrow}}}

\nc{\be}{\begin{equation}}
\nc{\ee}{{\end{equation}}}
\nc{\bea}{\begin{eqnarray}}
\nc{\eea}{\end{eqnarray}}
\nc{\<}{\langle}
\rnc{\>}{\rangle}
\nc{\Hom}[2]{\mbox{Hom}(\CC^{#1},\CC^{#2})}
\nc{\rU}{\mbox{U}}

\begin{document}

\author{
\authorblockN{Graeme Smith}
\authorblockA{IBM TJ Watson Research Center\\
1101 Kitchawan Road\\
 Yorktown NY 10598\\
gsbsmith@gmail.com}

}

\title{Quantum Channel Capacities}

\maketitle

\begin{abstract}
A quantum communication channel can be put to many uses: it can transmit classical information, private classical information, or quantum information.  It can 
be used alone, with shared entanglement, or together with other channels.  For each of these settings there is a capacity that quantifies a channel's potential
for communication.  In this short review, I summarize what is known about the various capacities of a quantum channel, including a discussion of the relevant 
additivity questions.  I also give some indication of potentially interesting directions for future research. 
\end{abstract}

\section{Introduction}

The capacity of a noisy communication channel for noiseless information transmission is 
a central quantity in the study of information theory \cite{CoverThomas}.  This capacity
establishes the ultimate boundary between communication rates which are achievable in principle and those which 
are not.  Furthermore, knowing a noisy channel's capacity can guide the design of explicit coding strategies
as well as giving us a benchmark for testing practical communication schemes.

The usual starting point for information theory is to model the communication channel stochastically.  We think 
of a channel as a noisy mapping of some input $x$ to an output $y$ according to some transition probabilities, $p(y|x)$.
Then, we can find a simple formula for a channel's capacity as a function of these parameters:  it's the maximum mutual 
information that can be generated between input and output given a single use of the channel.

However, this model is not rich enough to include quantum effects.  Therefore, it's necessary to develop a more general
information theory that takes quantum mechanics into account.  Quantum information theory addresses the question of noisy 
data transmission above, but also explores applications of quantum effects to other communication and cryptographic tasks.  As a result
a quantum channel has a variety of capacities, each of which characterize its capability for achieving a different kind of communication
task.  For example, the most direct analogue of the capacity mentioned above is the {\em classical capacity} of a quantum channel, which 
tells us the best rate at which the channel can transmit classical information from sender to receiver.  The {\em private classical capacity}
of a quantum channel quantifies its capability for quantum cryptography \cite{BB84}, and has close connections to the capacity of a wire-tap channel
as considered by Wyner \cite{Wyn75} and Csiszar-Korner \cite{CK78}.  If we are interested in coherently transmitting a quantum state, we must consider the 
{\em quantum capacity} of our channel, and if we have access to arbitrary quantum correlations between sender and receiver the relevant capacity is the
{\em entanglement assisted capacity}.  In general, these capacities are all different, which gives a variety of inequivalent ways to quantify 
the value of a quantum channel for communication.

The communication capacities of a quantum channel are not nearly as well understood as their classical counterparts, and many basic questions 
about quantum capacities remain open.  The purpose of this paper is to give an introduction to a quantum channel's capacities, summarize what we know 
about them, and point towards some important unsolved problems.

\section{Quantum States and Channels}

The states of least uncertainty in quantum mechanics are {\em pure states}.  A pure state of a $d$-level
quantum system is described by a unit vector in a $d$-dimensional complex vector space:  $\ket{\psi} \in \CC^d$ with 
$\braket{\psi}{\psi} = 1$.  In the simplest form of measurement on a quantum system (a ``von Neumann measurement''), 
the experimenter chooses an orthogonal basis, $\{\ket{i}\}_{i=1}^d$, for $\CC^d$.  The measurement ``projects'' the 
state $\ket{\psi}$ into outcome $i$ with probability $|\braket{i}{\psi}|^2$.  Thus, if a system is prepared in a basis 
state $\ket{i}$ and measured, the outcome is always $i$.  However, in general quantum mechanics will only tell us the 
probabilities of measurement outcomes. 

Pure states are the states of least uncertainty in quantum mechanics, but one could imagine situations that are 
more uncertain.  In such cases, rather than a $d$-dimensional complex vector, the state of our system is described
by a $d \times d$ hermitian matrix, $\rho = \rho^\dagger$.  Any such {\em density matrix} has a spectral decomposition, 
$\rho = \sum_i \lambda_i \proj{\phi_i}$, which can be interpreted as telling us that the system is in pure state $\ket{\phi_i}$
with probability $\lambda_i$.  In order to ensure this probabilistic interpretation makes sense, $\rho$ must be trace one: $\Tr \rho =1$.
Suppose we have a density matrix of a system comprising two subsystems, $A$ and $B$.  Given such a $\rho_{AB}$, we can find the density matrix of
the $A$ system alone using the partial trace, so that $\rho_A = \Tr_B \rho_{AB}$.  

The noisy evolution of a quantum system can be described as a unitary interaction between the system and some
environmental degrees of freedom.  More formally, any noisy evolution has the form ${\cal N}(\rho_A) = \Tr_E U \rho_A \otimes \proj{0}_E U^\dagger$,
where $\ket{0}$ is some fixed pure state on $E$ and $U$ is a unitary matrix on from $AE$ to $BE$.  We use $B$ to denote the output of the channel
to allow for the possibility that the input and output spaces are different.  Mathematically, such a map is called a completely positive 
trace preserving map, but I'll just call it a quantum channel.  There is an alternate representation of a quantum channel in terms of {\em Kraus operators}, 
which lets us write the channel's action as ${\cal N}(\rho) = \sum_k A_k \rho A_k^\dagger$ with $\sum_k A_k ^\dagger A_k = I$.

More details on the basics of quantum states and channels can be found in \cite{NC2000}.

\section{Coding Theorems for Quantum Channels}

I'll now discuss four fundamental coding theorems of quantum communications: coding for classical information transmission, private classical transmission,
quantum transmission, and entanglement assisted classical transmission.  I won't go into the proofs of these coding theorems but, roughly speaking, they 
all amount to randomly coding according to some distribution at the input of the channel.  Notions like typical sequences and conditionally typical sequences, 
suitably generalized to the quantum domain, are key ingredients. 

\subsection{Classical Transmission}

Suppose a sender is interested in using a noisy quantum channel to send classical information to a receiver.  The maximum rate at which this is possible, 
measured in bits per channel use and with the probability of a transmission error vanishing in the limit of many channel uses, is defined as the {\em classical
capacity} of a quantum channel, which we'll call $C({\cal N})$.

Holevo \cite{Holevo98a} and Schumacher and Westmoreland \cite{Schumacher97a} found a lower bound for $C({\cal N})$: 

\begin{equation}\label{Eq:C}
C({\cal N}) \geq \chi({\cal N}),
\end{equation}
where the {\em Holevo Information} \cite{Holevo73} is defined as 
$\chi({\cal N}) = \max_{p_x, \rho_x} I(X;B)_{\sigma}$, 
where $\sigma = \sum_x p_x \proj{x}_X \otimes {\cal N}(\rho_x)$.  The mutual information is defined as $I(X;B) = H(X) + H(B) - H(XB)$, 
with the von Neuman entropy, $H(\rho) = -\Tr\rho \log \rho$, replacing the usual Shannon entropy.

In addition to showing that $C({\cal N}) \geq \chi({\cal N})$, it's also possible to give a characterization of the classical capacity as 
$C({\cal N}) = \lim_{n \rightarrow \infty}\frac{1}{n}\chi({\cal N}^{\otimes n})$ \cite{Holevo98a,Schumacher97a}.  Of course this limit is not something that can actually be computed.  

\subsection{Private Classical Transmission}

Now suppose we have access to a noisy quantum channel, and we would again like to transmit classical information from sender to receiver.  This time, however,
we're interested in making sure that {\em only} the receiver learns what message we sent.  Specifically, we have to make sure that no information about the 
transmitted message gets leaked to the environment of the channel.  This situation is qualitatively similar to the wire-tap channel considered by Wyner \cite{Wyn75} and Csiszar-Korner \cite{CK78} and, in fact, the solution looks similar in many ways.

The resulting {\em private classical capacity}, roughly the best rate that we can achieve with both high fidelity for the receiver and no information leaked to the
environment, is usually called $P({\cal N})$.  Devetak \cite{D03} and Cai-Winter-Yeung \cite{CWY04} showed that the
private classical capacity satisfies

\begin{equation}
P({\cal N}) \geq P^{(1)}({\cal N}) := \max_{p_x, \rho_x} I(X;B)_\sigma - I(X;E)_\sigma,
\end{equation}
where $\sigma = \sum_x p_x \proj{x}_X \otimes U ( \rho_x \otimes \proj{0}_E ) U^\dagger$ and $U$ is the unitary extension of
the channel $\cal N$.  The expression on the right is very similar to the capacity formula found in a classical broadcast
setting \cite{CK78}.

Just as for the classical capacity of a quantum channel, it can also be shown that the private classical capacity
satisfies $P({\cal N}) = \lim_{n \rightarrow \infty}\frac{1}{n}P^{(1)}({\cal N}^{\otimes n})$.

\subsection{Quantum Transmission}
If our sender has a quantum system whose state she would like to transmit coherently to a receiver, we enter the realm of the 
quantum capacity.  We'd like to understand the best rate, this time in two-level quantum systems (or {\em qubits}) per channel use, at which 
we can coherently transmit quantum information.  This requires that an arbitrary quantum state, when encoded and transmitted using a noisy channel, can 
be recovered by the receiver.  As a result the quantum capacity characterizes the fundamental limits of {\em quantum error correction}.

A lower bound for the quantum capacity, usually called $Q({\cal N})$, was found by Lloyd \cite{Lloyd97}, Shor \cite{Shor02}, and Devetak \cite{D03}.  They 
found that the {\em coherent information} is an achievable rate:
\begin{equation}
Q({\cal N}) \geq Q^{(1)}({\cal N}):= \max_{\rho} \left( H(B) - H(E)\right),
\end{equation}
where the entropies are evaluated on the reduced states of $\sigma_{BE}= U \rho \otimes \proj{0}_E U^\dagger$.  The coherent information can also be expressed
as a maximization of $-H(A|B)= H(B)- H(AB)$, evaluated on a state of the form $I \otimes {\cal N}(\proj{\phi}_{AA'})$, but I won't use this fact here.

Finally, just like with the classical and private capacities, we can get a characterization of the quantum capacity as
 $Q({\cal N}) = \lim_{n \rightarrow }\frac{1}{n}Q^{(1)}({\cal N}^{\otimes n})$.

\subsection{Classical Transmission with Entanglement Assistance}
At this point let's go back to trying to understand the transmission of classical information with a quantum channel.  However, besides letting our sender and 
receiver use a noisy quantum channel, we'll also give them access to arbitrary shared quantum states  (and specifically, they'll use {\em entangled} states).  
On their own, such quantum states are useless for communication due to the locality of quantum mechanics.  However, the shared quantum states 
may in principle be useful when used together with a noisy channel.  In fact, not only will the shared quantum correlations be useful for assisting many quantum 
channels, it turns out that they will lead to a dramatic simplification of the theory.

Define the {\em entanglement assisted classical capacity} to be the maximum rate at which classical communication is possible with low error probability
when sender and receiver use a noisy channel together with arbitrary shared quantum states.  Bennett, Shor, Smolin, and Thapliyal \cite{BSST02} showed that
this capacity is given by 

\begin{equation}\label{Eq:CE}
C_E({\cal N}) = \max_{\phi_{AA'}}I(A;B)_{\sigma},
\end{equation}
where the quantum mutual information is evaluated on the state $\sigma = I \otimes {\cal N}({\phi}_{AA'})$.  Note that the only 
difference between Eq.~(\ref{Eq:CE}) and the Holevo information of Eq.~(\ref{Eq:C}) is that the Holevo information is the maximum mutual information
we can generate using a state of the form $\phi_{XA} = \sum_x p_x \proj{x} \otimes \rho_x$, whereas in Eq.~(\ref{Eq:CE}) the state $\phi_{AA'}$ is unrestricted.

The truly remarkable thing about Eq.~(\ref{Eq:CE}) is that we have an equality---the formula is single-letter.  There is no need to take a limit over many 
channels uses, so this formula gives a complete characterization of the channel's capability for classical transmission given free access to entanglement.
The appearance of a single letter formula here, and the absence of single-letter formulas for the three other capacities we've discussed is closely related
to the question of additivity of information theoretic quantities, which I'll consider in the next section.

\section{Additivity Questions}
A real function, $f$, on the set of channels is called additive if
$f({\cal N}\otimes {\cal M}) = f({\cal N}) + f({\cal M})$.  Many important questions in quantum information theory boil down to asking 
whether a certain function is additive.  Very often, we have some function that it's fairly easy to see satisfies 
$f({\cal N}\otimes {\cal M}) \geq f({\cal N}) + f({\cal M})$, but we would like to show equality.

For example, the single-letter formula for the entanglement assisted classical capacity is shown in three steps.  
Let $g({\cal N}) = \max_{\phi_{AA'}} I(A;B)_{\sigma}$ with $\sigma_{AB} = I\otimes {\cal N}(\phi_{AA'})$.  The first step is to show that $C_E({\cal N}) \geq g({\cal N})$ 
by a random coding argument.  The next step is to show that $C_E({\cal N}) \leq \lim_{n \rightarrow \infty}\frac{1}{n}g({\cal N}^{\otimes n})$, basically by using
the continuity properties of von Neuman entropy together with the requirement for asymptotically near-perfect transmission.  Finally, one shows (in this
case, using strong subadditivity of entropy) that $g$ is additive, from which we get 
$C_E({\cal N}) \leq \lim_{n \rightarrow \infty}\frac{1}{n}g({\cal N}^{\otimes n}) = g({\cal N})$, which gives $C_E({\cal N}) = g({\cal N})$.

Conversely, often where we would have liked to show additivity it turns out not to be true.  One of the earliest 
examples of this was the finding of \cite{DSS98} that the coherent information is not additive on several copies of 
a very noisy qubit depolarizing channel\footnote{The qubit depolarizing channel is the most natural quantum analogue of a binary symmetric channel.  
It acts as ${\cal N}_p(\rho) = (1-p)\rho + p \frac{I}{2}$.}. Specifically, they showed that for a range of depolarizing parameter $p$, 
$Q^{(1)}({\cal N}_p^{\otimes 5})> 5 Q^{(1)}({\cal N}_p)$, thus dashing hopes of proving $Q({\cal N}) = Q^{(1)}({\cal N})$.  More recently, it was shown 
in \cite{SRS08} that the private information, $P^{(1)}$, is not additive on several copies of a channel closely related to the Bennett-Brassard-84 quantum cryptography 
protocol \cite{BB84}.  Finally, Hastings recently showed that there are channels such that $\chi({\cal N} \otimes {\cal N}) > 2 \chi({\cal N})$.  In each of these
cases---the nonadditivity of $Q^{(1)}$, $P^{(1)}$, and $\chi$---there was a natural guess for a simple formula giving the associated capacity, but the nonadditivity 
implied that the capacity did not equal the formula.  This leaves us without an effectively computable characterization of the capacities, 
but in each case we also find that the natural guess for capacity was overly pessimistic---by using more complicated coding strategies we can transmit more information.

Although $\chi$, $P^{(1)}$, and $Q^{(1)}$ are not additive in general, it is possible to show additivity for some special classes of channels.  For example, 
$C({\cal N}) = \chi({\cal N})$ for all entanglement breaking channels\footnote{Entanglement breaking channels are so noisy that they don't allow sender and receiver
to establish any entanglement.  Alternatively, an entanglement breaking channel has a set of rank one Kraus operators.} \cite{ShorEB02}, 
depolarizing channels \cite{King03}, and unital qubit channels.  The quantum capacity is equal to $Q^{(1)}({\cal N})$ 
for both
degradable channels \cite{DS03} and PPT channels \cite{HHH96,Peres96} although for the latter it is 
always zero.  Finally, the private capacity is $P^{(1)}({\cal N})$ for degradable channels \cite{Smith08}.

The additivity questions considered so far concern the need for {\em regularization} 
(as taking $\lim_{n \rightarrow \infty} \frac{1}{n}f({\cal N}^{\otimes n})$ is called) in our characterization of capacities.
Some entropic function $f({\cal N})$ is shown to be an achievable rate, and its regularization equal to the capacity, so 
if we can show $f$ is additive we get a tractable capacity formula.  Note that this concerns the additivity of $f$ on
two copies of the {\em same} channel.  However, once we regularize $f$  (consider, say, 
the formula $Q({\cal N}) = \lim_{n \rightarrow \infty}\frac{1}{n} Q^{(1)}({\cal N}^{\otimes n})$), the resulting capacity always 
satisfies $Q({\cal N}^{\otimes n}) = nQ({\cal N})$.  Capacities are always additive on parallel uses of the same channel.

The need for regularization is a mathematical question about the existence of a simple formula for capacity, but there 
is a distinct, more operational, question about the additivity of capacities themselves.  Since capacities are additive
on products of the same channel, this is a question about how different noisy channels interact and enhance each others' 
capabilities\footnote{Note that it is possible to have a nonadditive $f$ that, when regularized, gives an additive
capacity.  Similarly, one could have a simple formula for a nonadditive capacity.  So, one type of (non)additivity 
does not generally imply the other.}.  This sort of additivity thus tells us whether the communication potential of 
a channel depends on the context in which it is used or is independent of what other channels are available with it.  The
quantum capacity is very strongly nonadditive:  there are pairs of 
channels ${\cal N}$ and ${\cal M}$ with $Q({\cal N}) = Q({\cal M}) = 0$ but $Q({\cal N}\otimes{\cal M})>0$ \cite{SY08}.
The private capacity displays nonadditivity that is almost as strong, with 
$P({\cal N}) = 0$ and $P({\cal M}) \leq 2$  but $P({\cal N}\otimes{\cal M})\geq 1/8 \log d$ for sufficiently large
dimensional channels \cite{SS09a,KeLi09,SS09b}.  It is unknown whether the classical capacity of a quantum channel 
is additive in this sense.

\begin{figure}
\begin{tabular}{|c|c|c|}
\hline
{\tiny Information} $\backslash$ {\tiny Additivity} & Capacity & Single-letter\\ 
\hline
Classical & ? & No \cite{Hastings09}\\
\hline
Private & No \cite{SS09a,KeLi09,SS09b} & No \cite{SRS08}\\
\hline
Quantum & No \cite{SY08}  & No \cite{DSS98}\\
\hline
Entanglement  & Yes \cite{BSST02} & Yes \cite{BSST02}\\
Assisted & & \\
\hline
\end{tabular}

\caption{Status of additivity questions for quantum capacities.  The different rows correspond to using the channel for different kinds of information
transmission.  The right column indicates whether regularization is necessary in our characterization of the associated capacity.  The left column indicates
whether the capacity itself is additive.  Note that the right column concerns the additivity of entropic quantities when evaluated on multiple copies of {\em the same}
channel, while the left concerns the behavior of a capacity when evaluated on two copies of {\em different} channels.}
\end{figure}

\section{Outlook}

There has been an enormous amount of progress on understanding the various communication capacities of a quantum
channel in the past decade 
or so.  From 1997-2003, the basic tools necessary to understand random coding in quantum communication were developed and,
more recently, much more streamlined approaches have been found \cite{Yard07}.  
In the past couple of years several fundamental questions about the additivity of capacities have been resolved, 
with the general trend that most things are not additive.

At first glance this nonadditivity seems like bad news.  The need for regularization means that we don't have simple
capacity formulas available, and the nonadditivity of capacities implies that the capacity of a channel might not even
be the relevant measure of its communication capability.  However, in both cases there is a positive side---regularization 
means that the capacities we're interested in are generally higher than expected, and we can use communication
strategies like entangled signal states and structured error correction codes to enhance our communication abilities.    
Nonadditivity of capacities means that even apparently useless resources may interact synergistically to allow 
communication and error correction where it appeared impossible. 
 
There are at least two interesting directions to pursue that are related to the classical capacity.  First, there
are very few computable upper bounds for the classical capacity.  The log of the input and output dimensions of a channel
are obviously bounds, as is the entanglement assisted capacity.  Shor also presented a bound of $\chi({\cal N})+ \max_{\rho_{AB}}E_F(I\otimes {\cal N}(\rho_{AB}))$, 
where $E_F$ is the entanglement of formation \cite{ShorEB02}.  None of these are particularly good in the low noise regime.  Second, although $\chi$ is not additive, 
we still don't know whether its regularization, $C({\cal N})$, is additive.

The bosonic gaussian channels are an important, potentially experimentally relevant, class 
that it would be nice to understand better (for a review, see \cite{WE05}).  Capacities of the 
attenuation channel are known \cite{Lossy04}, but even for single mode gaussian channels we only understand a few special cases (see \cite{Gio10} for recent
work on the classical capacity).  Since gaussian channels are a fairly simple class, understanding their capacities may be easier than the general case.  
As usual, the question hinges on understanding additivity properties of $\chi$, $P^{(1)}$, and $Q^{(1)}$.

In general, we need more examples of both additivity and nonadditivity if we are to understand coding for quantum channels.  For classical transmission, we 
understand the additivity of $\chi$ for entanglement breaking channels by a general argument.  There are also ad hoc techniques for understanding very specific
channels \cite{King03,EW05,Lossy04,BDS97}.  In terms of nonadditivity of $\chi$, there remains essentially a single class of nonconstructive 
examples \cite{Hastings09} (though, this class has been better understood recently \cite{Hast1, Hast2, Hast3}).  There are a few examples of nonadditivity
for $Q^{(1)}$, $P^{(1)}$, $Q$, and $P$ \cite{DSS98,SRS08,SY08,KeLi09,SS09b}, but it is still not entirely clear when nonadditivity should be expected (though
see \cite{Brand10} for some ideas).  For the quantum and private capacity, essentially the
only additivity results we have concern degradable channels, with PPT channels also understood for the quantum capacity.  It would be fantastic to find 
the quantum capacity of some channels that are not related to the degradable channels.

There is a qualitative similarity between quantum channels and classical broadcast channels \cite{CoverReview98}---in point-to-point quantum problems,
we always implicitly have a second receiver in the form of the environment.  So, when we look at the private classical transmission with a quantum channel, 
the achievable rates we find are closely related to the classical results on broadcasting privacy \cite{CK78}.  Furthermore, most of the channels whose quantum
capacity we can compute are degradable, a notion that comes directly from the classical idea of a degraded broadcast channel \cite{Cover72,CoverReview98}.  Can this correspondence be
developed into an explicit mapping between (perhaps a subset of) quantum channels and classical broadcast channels, with the capacities of the former derived
from the capacity region of the latter?

I'm grateful to Charlie Bennett for comments an an earlier draft and acknowledge support from DARPA QUEST contract HR0011-09-C-0047.

\begin{thebibliography}{10}
\providecommand{\url}[1]{#1}
\csname url@samestyle\endcsname
\providecommand{\newblock}{\relax}
\providecommand{\bibinfo}[2]{#2}
\providecommand{\BIBentrySTDinterwordspacing}{\spaceskip=0pt\relax}
\providecommand{\BIBentryALTinterwordstretchfactor}{4}
\providecommand{\BIBentryALTinterwordspacing}{\spaceskip=\fontdimen2\font plus
\BIBentryALTinterwordstretchfactor\fontdimen3\font minus
  \fontdimen4\font\relax}
\providecommand{\BIBforeignlanguage}[2]{{%
\expandafter\ifx\csname l@#1\endcsname\relax
\typeout{** WARNING: IEEEtran.bst: No hyphenation pattern has been}%
\typeout{** loaded for the language `#1'. Using the pattern for}%
\typeout{** the default language instead.}%
\else
\language=\csname l@#1\endcsname
\fi
#2}}
\providecommand{\BIBdecl}{\relax}
\BIBdecl

\bibitem{CoverThomas}
T.~M. Cover and J.~A. Thomas, \emph{{E}lements of {I}nformation
  {T}heory}.\hskip 1em plus 0.5em minus 0.4em\relax Wiley \&{} {S}ons, 1991.

\bibitem{BB84}
C.~H. Bennett and G.~Brassard, ``Quantum cryptography: Public key distribution
  and coin tossing,'' \emph{Proceedings of the IEEE International Conference on
  Computers, Systems and Signal Processing}, p. 175, 1984.

\bibitem{Wyn75}
A.~D. Wyner, ``{The Wire-tap Channel},'' \emph{Bell Systems Technical Journal},
  vol.~54, no.~8, pp. 1355--1387, January 1975.

\bibitem{CK78}
I.~Csiszar and J.~Korner, ``Broadcast channels with confidential messages,''
  \emph{IEEE Trans. Inf. Theory}, vol.~24, pp. 339--348, 1978.

\bibitem{NC2000}
M.~A. Nielsen and I.~L. Chuang, \emph{Quantum Computation and Quantum
  Information}, 1st~ed.\hskip 1em plus 0.5em minus 0.4em\relax Cambridge
  University Press, October 2000.

\bibitem{Holevo98a}
A.~S. Holevo, ``The capacity of the quantum channel with general signal
  states,'' \emph{IEEE. Trans. Inf. Theory}, vol.~44, no.~1, pp. 269--273,
  1998.

\bibitem{Schumacher97a}
B.~Schumacher and M.~D. Westmoreland, ``Sending classical information via noisy
  quantum channels,'' \emph{Phys. Rev. A}, vol.~56, no.~1, pp. 131--138, 1997.

\bibitem{Holevo73}
A.~S. Holevo, ``Statistical problems in quantum physics,'' in \emph{Proceedings
  of the second {Japan-USSR} Symposium on Probability Theory}, ser. Lecture
  Notes in Mathematics, G.~Maruyama and J.~V. Prokhorov, Eds., vol. 330.\hskip
  1em plus 0.5em minus 0.4em\relax Berlin: Springer-Verlag, 1973, pp. 104--119.

\bibitem{D03}
I.~Devetak, ``The private classical capacity and quantum capacity of a quantum
  channel,'' \emph{IEEE Trans. Inf. Theory}, vol.~51, pp. 44--55, 2005,
  ar{X}iv:quant-ph/0304127.

\bibitem{CWY04}
N.~Cai, A.~Winter, and R.~Yeung, ``Quantum privacy and quantum wiretap
  channels,'' \emph{Problems of Information Transmission}, vol.~40, no.~4, pp.
  318--336, 2004.

\bibitem{Lloyd97}
S.~Lloyd, ``Capacity of the noisy quantum channel,'' \emph{Phys. Rev. A},
  vol.~55, pp. 1613--1622, 1997.

\bibitem{Shor02}
P.~W. Shor, ``The quantum channel capacity and coherent information.'' lecture
  notes, MSRI Workshop on Quantum Computation, 2002. Available online at
  http://www.msri.org/publications/ln/msri/2002/\\ quantumcrypto/shor/1/.

\bibitem{BSST02}
C.~H. Bennett, P.~W. Shor, J.~A. Smolin, and A.~V. Thapliyal,
  ``Entanglement-assisted capacity of a quantum channel and the reverse shannon
  theorem,'' \emph{IEEE Trans. Inf. Theory}, vol.~48, pp. 2637--2655, 2002.

\bibitem{DSS98}
D.~Di{V}incenzo, P.~W. Shor, and J.~A. Smolin, ``Quantum channel capacity of
  very noisy channels,'' \emph{Phys. Rev. A}, vol.~57, no.~2, pp. 830--839,
  1998, ar{X}iv:quant-ph/9706061.

\bibitem{SRS08}
G.~{Smith}, J.~M. {Renes}, and J.~A. {Smolin}, ``{Structured Codes Improve the
  Bennett-Brassard-84 Quantum Key Rate},'' \emph{Physical Review Letters}, vol.
  100, no.~17, p. 170502, May 2008, arXiv:quant-ph/0607018.

\bibitem{ShorEB02}
P.~W. {Shor}, ``{Additivity of the classical capacity of entanglement-breaking
  quantum channels},'' \emph{Journal of Mathematical Physics}, vol.~43, pp.
  4334--4340, Sep. 2002, arXiv:quant-ph/0201149.

\bibitem{King03}
C.~King, ``The capacity of the quantum depolarizing channel,'' \emph{IEEE
  Trans. Info. Theo.}, vol.~49, no.~1, pp. 221--229, 2003.

\bibitem{DS03}
I.~Devetak and P.~W. Shor, ``The capacity of a quantum channel for simultaneous
  transmission of classical and quantum information,'' \emph{Comm. Math.
  Phys.}, vol. 256, no.~2, pp. 287--303, 2005, ar{X}iv:quant-ph/0311131.

\bibitem{HHH96}
M.~Horodecki, P.~Horodecki, and R.~Horodecki, ``Separability of mixed states:
  necessary and sufficient conditions,'' \emph{Phys. Lett. A}, vol. 223, no.
  1-2, pp. 1--8, 1996.

\bibitem{Peres96}
A.~Peres, ``Separability criterion for density matrices,'' \emph{Phys. Rev.
  Lett.}, vol.~77, pp. 1413--1415, 1996.

\bibitem{Smith08}
G.~{Smith}, ``Private classical capacity with a symmetric side channel and its
  application to quantum cryptography,'' \emph{Phys. Rev. A}, vol.~78, no.~2,
  p. 022306, 2008, ar{X}iv:0705.3838.

\bibitem{SY08}
G.~{Smith} and J.~{Yard}, ``{Quantum Communication with Zero-Capacity
  Channels},'' \emph{Science}, vol. 321, pp. 1812--1816, Sep. 2008,
  ar{X}iv:0807.4935.

\bibitem{SS09a}
G.~{Smith} and J.~A. {Smolin}, ``{Can Nonprivate Channels Transmit Quantum
  Information?}'' \emph{Physical Review Letters}, vol. 102, no.~1, p. 010501,
  Jan. 2009, ar{X}iv:0810.0276.

\bibitem{KeLi09}
K.~{Li}, A.~{Winter}, X.~{Zou}, and G.~{Guo}, ``{Private Capacity of Quantum
  Channels is Not Additive},'' \emph{Physical Review Letters}, vol. 103,
  no.~12, p. 120501, Sep. 2009, ar{X}iv:0903.4308.

\bibitem{SS09b}
G.~{Smith} and J.~A. {Smolin}, ``{Extensive Nonadditivity of Privacy},''
  \emph{Physical Review Letters}, vol. 103, no.~12, p. 120503, Sep. 2009,
  ar{X}iv:0904.4050.

\bibitem{Hastings09}
M.~B. {Hastings}, ``{Superadditivity of communication capacity using entangled
  inputs},'' \emph{Nature Physics}, vol.~5, pp. 255--257, Apr. 2009,
  ar{X}iv:0809.3972.

\bibitem{Yard07}
P.~{Hayden}, M.~{Horodecki}, J.~{Yard}, and A.~{Winter}, ``{A decoupling
  approach to the quantum capacity},'' \emph{Open Syst. Inf. Dyn.}, vol.~15,
  pp. 7--19, 2008, ar{X}iv:quant-ph/0702005.

\bibitem{WE05}
J.~{Eisert} and M.~M. {Wolf}, ``{Gaussian quantum channels},'' 2005,
  ar{X}iv:quant-ph/0505151.

\bibitem{Lossy04}
V.~{Giovannetti}, S.~{Guha}, S.~{Lloyd}, L.~{Maccone}, J.~H. {Shapiro}, and
  H.~P. {Yuen}, ``{Classical Capacity of the Lossy Bosonic Channel: The Exact
  Solution},'' \emph{Physical Review Letters}, vol.~92, no.~2, p. 027902, 2004,
  ar{X}iv:quant-ph/0308012.

\bibitem{Gio10}
V.~{Giovannetti}, A.~S. {Holevo}, S.~{Lloyd}, and L.~{Maccone}, ``{Generalized
  minimal output entropy conjecture for Gaussian channels: definitions and some
  exact results},'' 2010, ar{X}iv:1004.4787.

\bibitem{EW05}
M.~M. {Wolf} and J.~{Eisert}, ``{Classical information capacity of a class of
  quantum channels},'' \emph{New Journal of Physics}, vol.~7, p.~93, Apr. 2005,
  ar{X}iv:quant-ph/0412133.

\bibitem{BDS97}
C.~H. Bennett, D.~P. DiVincenzo, and J.~A. Smolin, ``Capacities of quantum
  erasure channels,'' \emph{Phys. Rev. Lett.}, vol.~78, no.~16, pp. 3217--3220,
  Apr 1997.

\bibitem{Hast1}
F.~G.~S.~L. {Brandao} and M.~{Horodecki}, ``{On Hastings' counterexamples to
  the minimum output entropy additivity conjecture},'' Jul. 2009,
  ar{X}iv:0907.3210.

\bibitem{Hast2}
M.~{Fukuda}, C.~{King}, and D.~K. {Moser}, ``{Comments on Hastings' Additivity
  Counterexamples},'' \emph{Communications in Mathematical Physics}, vol. 296,
  pp. 111--143, May 2010, ar{X}iv:0905.3697.

\bibitem{Hast3}
G.~{Aubrun}, S.~{Szarek}, and E.~{Werner}, ``{Hastings' additivity
  counterexample via Dvoretzky's theorem},'' \emph{ArXiv e-prints}, Mar. 2010,
  ar{X}iv:1003.4925.

\bibitem{Brand10}
F.~G.~S.~L. {Brand{\~a}o} and J.~{Oppenheim}, ``{Public Quantum Communication
  and Superactivation},'' 2010, ar{X}iv:1005.1975.

\bibitem{CoverReview98}
T.~Cover, ``Comments on broadcast channel,'' \emph{IEEE Trans. Info. Theo.},
  vol.~44, no.~6, pp. 2524--2529, 1998.

\bibitem{Cover72}
------, ``Broadcast channels,'' \emph{IEEE Trans. Inf. Theory}, vol.~18, pp.
  2--14, 1972.

\end{thebibliography}


\end{document}